\documentclass[useAMS,usenatbib]{mn2e}
\usepackage{amsfonts}
\usepackage{amsmath}
\usepackage{amssymb}
\usepackage{graphicx}
\usepackage{float}

\newcommand{\kps}{\rm{\,km\,s}^{-1}}
\newcommand{\mG}{~\umu \rm{G}}

\newcommand{\highlight}[1]{#1} %Uncomment for no highlight version

\title[Fast and slow MHD shock waves]{Signatures of fast and slow magnetohydrodynamic shocks in turbulent molecular clouds}
\author[Andrew Lehmann and Mark Wardle]{Andrew Lehmann\thanks{E-mail:
andrew.lehmann@mq.edu.au} and Mark Wardle\\
Department of Physics and Astronomy, and Research Centre for Astronomy, Astrophysics \& Astrophotonics,\\
Macquarie University, Sydney, NSW 2109, Australia}

\begin{document}

\pagerange{\pageref{firstpage}--\pageref{lastpage}} \pubyear{2015}

\maketitle

\begin{abstract}

The character of star formation is intimately related to the supersonic magnetohydrodynamic (MHD) turbulent dynamics of the molecular clouds in which stars form. A significant amount of the turbulent energy dissipates in low velocity shocks. Fast and slow MHD shocks differ in how they compress and heat the molecular gas, and so their radiative signatures reveal distinct physical conditions.

We use a two-fluid model to compare one-dimensional fast and slow MHD shocks propagating at low speeds (a few $\kps$). Fast shocks are magnetically driven, forcing ion species to stream through the neutral gas ahead of the shock front. This magnetic precursor heats the gas sufficiently to create a large, warm transition zone where all the fluid variables smoothly change in the shock front. In contrast, slow shocks are driven by gas pressure, and neutral species collide with ion species in a thin hot slab that closely resembles an ordinary gas dynamic shock. 

We consider shocks at velocities $v_s = 2$--$4 \kps$ and preshock Hydrogen nuclei densities $n_\mathrm{H} = 10^2$--$10^4$ cm$^{-3}$. We include a simple oxygen chemistry and cooling by CO, H$_2$ and H$_2$O. CO rotational lines above $J = 6\to 5$ are more strongly excited in slow shocks. These slow shock signatures may have already been observed in infrared dark clouds in the Milky Way.

\end{abstract}


\section{Introduction} \label{ch:Intro}

Understanding the internal environment of the giant molecular clouds (GMCs) in which stars form is a necessary precursor to addressing the star formation rate and stellar initial mass function \citep[e.g.,][]{bergin_cold_2007,mckee_theory_2007}. \highlight{Large non-thermal linewidths observed in molecular lines \citep[e.g.,][]{larson_turbulence_1981,solomon_mass_1987} have often been attributed to turbulent motions. Furthermore, the kinetic energy associated with these motions is generally found to be on the order of the gravitational potential energy, indicating the importance of turbulence as a dynamical component of molecular clouds. In addition, magnetic fields in GMCs give Alfv\'{e}n velocities on the order of the observed velocity dispersions \citep[e.g.,][]{crutcher_oh_1993,crutcher_magnetic_1999, crutcher_magnetic_2010}.} This suggests that the turbulence in GMCs is magnetohydrodynamic (MHD) in nature. 

\highlight{The physics underlying the MHD turbulence of molecular clouds is intimately connected with properties of star formation \citep{mac_low_control_2004}. For example, \citet{federrath_star_2012} and \citet{federrath_star_2013} use three dimensional MHD simulations to analyse different modes of turbulence by comparing compressive driving to solenoidal driving. They showed that the star formation rate and efficiency are both sensitive to these driving modes and Mach number variations. Hence it is desirable to discover observable distinctions between different modes of turbulence.}

\highlight{Supersonic turbulence dissipates via shock waves and vortices \citep{pety_elusive_2000}. In simulations of compressible MHD turbulence \citet{stone_dissipation_1998} found that shock waves dissipated 50\% of the turbulent energy in a strong magnetic field model and 32\% in a weak field model. The heating of compressed gas in a thin postshock region uniquely drives chemistry and radiative cooling. Thus the radiative characteristics of turbulent dissipation will be strongly shaped by cooling in shocks. Furthermore, \citet{smith_shock_2000} found that weak shocks in a large range of velocities were responsible for the majority of dissipation in simulations of decaying MHD turbulence. In contrast, a small range of stronger shocks dissipated turbulence that was being driven \citep{smith_distribution_2000}. Hence radiative signatures of shocks could be used to distinguish between these two scenarios.}

The observational signatures of the low-velocity shocks that dominate dissipation of MHD turbulence have only recently been considered. \citet{pon_molecular_2012} considered C-type fast MHD shocks travelling at speeds of $2$--$3 \kps$ perpendicular to the magnetic field and computed the abundances and emission of H$_2$, CO and H$_2$O. By comparing CO rotational line emission from these shocks to those produced in photodissociation regions (PDRs), they found that fast shocks dominate the emission in transitions above $J = 5 \to 4$. \citet{lesaffre_low-velocity_2013} take a statistical approach by computing observational diagnostics due to a distribution of C- and J-type fast, perpendicular, MHD shocks at velocities ranging from $3$--$40 \kps$. They use these shocks to explain the radiation from a turbulent wake formed by a galaxy collision in Stephan's Quintet.

Anomalously bright CO lines above $J = 5 \to 4$ have been recently observed towards Milky Way molecular clouds \citep[e.g.,][]{pon_mid-j_2014, pon_mid-j_2015, larson_evidence_2015} and from warm molecular gas in external galaxies \citep[e.g.,][]{kamenetzky_herschel-spire_2012, pellegrini_shock_2013}. These studies all conclude that PDR models are unable to reproduce the bright high-$J$ CO lines and all suggest MHD shock waves as the heating mechanism. \highlight{The shock models used or referred to in these studies \citep{flower_excitation_2010, pon_molecular_2012} and other studies of shocks in interstellar clouds \citep[e.g.][]{draine_magnetohydrodynamic_1986, hollenbach_molecule_1989, chapman_dust_2006} all consider \textit{fast} MHD shocks. MHD fluids can, however, support three kinds of shocks: fast, intermediate and slow. Unfortunately no study has identified which kinds of shocks would be produced by MHD turbulence. One of the goals of our work is to motivate the classification of MHD shocks in turbulent molecular clouds.}

\highlight{Ideal MHD assumes a fully ionized gas in which the magnetic field is frozen. Molecular clouds are in fact only weakly ionized and the magnetic field acts on the neutral fluid via the ionized fluid. A two-fluid description is therefore more appropriate to molecular cloud studies. Early work by \cite{lithwick_compressible_2001} and \cite{lazarian_magnetic_2004} highlighted the importance of two-fluid effects on the turbulent cascade. Recently it has become feasible to run high resolution three dimensional two-fluid MHD simulations. For example, \cite{meyer_observational_2014} use simulations to show that observations of linewidth differences between line emission from neutral and ion species can be accounted for by two-fluid effects. Furthermore, \cite{burkhart_alfvenic_2015} use simulations to show that the Alfv\`{e}nic modes do not necessarily dissipate at the ambipolar diffusion scale. They also confirm the analytic results of \cite{balsara_wave_1996} and \cite{tilley_two-fluid_2011} that in molecular cloud conditions, the fast and Alfv\`{e}n modes are strongly damped by ion-neutral collisions leaving the slow modes to propagate with little damping. Thus nonlinear steepening into slow shocks might be expected to preferentially occur in molecular clouds. In addition, slow shocks reach far higher peak temperatures than fast shocks for two reasons. Firstly, the heating timescale in fast shocks is determined by the long ion-neutral collision timescale while in slow shocks is determined by the short neutral-ion collision timescale. As the cooling timescale of the gas lies between these two, the gas being overrun by fast shocks remains at low temperatures whereas it quickly heats up within slow shocks. Secondly, as we show in Section~\ref{sec:Theory} the gas in fast shocks necessarily loses some of its kinetic energy to strengthening the magnetic field while in slow shocks it does not. Thus slow shocks have more energy available to heat the gas.}

\highlight{In this paper, we solve the steady, plane parallel two-fluid MHD equations to model shocks that propagate at any angle to the magnetic field. In Section~\ref{sec:Theory} we elucidate some of the basic differences of fast and slow MHD shocks. We describe our computational scheme in Section~\ref{ch:Integration}. In Section~\ref{sec:Results} we compare fast and slow shocks in the low-velocity regime with molecular cloud conditions. We discuss their radiative characteristics and implications for interpreting emission from turbulent molecular gas. Finally, in Section~\ref{sec:Discussion} we show how these signatures might be used to interpret observations.}

\section{Theory}\label{sec:Theory}

Cosmic rays streaming through the interstellar medium weakly ionize molecular clouds, generating an ion fluid that interpenetrates the neutral particles. A two-fluid MHD description is developed in Section~\ref{sec:MHD} with the ion fluid coupled to the magnetic field via the Lorentz force and to the neutral fluid via the collisional force, so that
\begin{align*}
\frac{\textbf{J} \times \textbf{B}}{c} = \alpha \rho _i \rho _n \left( \textbf{v}_i - \textbf{v}_n \right)
\end{align*}
where $\alpha$ is the rate coefficient for elastic ion-neutral scattering and $\rho$ and $\textbf{v}$ are the density and velocity with subscripts $i$ and $n$ referring to ion and neutral fluids respectively. By considering the state of these fluids far away from the shock front, we illustrate how the various families of MHD shocks come about and highlight the different effects they have on the ambient magnetic field.

A simplified chemical model, in which two body reactions and photodissociation affect the abundances of coolants, is presented in Section~\ref{sec:Chemistry}. Finally, the details of how those coolants radiate are outlined in Section~\ref{sec:Cooling}. Only rotational excitations of molecular species are considered because vibrational excitations do not become significant until temperatures reach $\gtrsim 1000$ K, which are not achieved in the low velocity shocks computed here.

\subsection{Two-Fluid MHD} \label{sec:MHD}
\subsubsection*{Conservation equations}
\highlight{While turbulence is an inherently three-dimensional problem \citep[e.g.][and references therin]{burkhart_alfvenic_2015}, shock waves are extremely thin structures within a turbulent system. Hence we follow \cite{draine_multicomponent_1986} by considering} a stationary plane-parallel shock travelling in the $z$ direction with the magnetic field initially lying in the $x$--$z$ plane. The governing equations of the ion fluid are
\begin{align}
& \frac{d}{dz} \left(\rho _{i} v_{iz} \right) = 0, \label{eq:ion_continuity}\\
& \frac{d}{dz} \left( \frac{B_z B_x }{4 \pi} \right) =  \alpha \rho _i \rho _n \left( v_{ix} - v_{nx} \right), \label{eq:ion_momentumx}\\
& \frac{d}{dz} \left( \frac{B_x^2}{8 \pi} \right) = -\alpha \rho _i \rho _n \left( v_{iz} - v_{nz} \right). \label{eq:ion_pressureB}
\end{align}
The neutral fluid equations are
\begin{align}
& \frac{d}{dz} \left( \rho _n  v_{nz} \right) = 0 \label{eq:neutral_continuity},\\
& \frac{d}{dz} \left( \rho _n v_{nz} v_{nx}\right)  = \alpha \rho _i \rho _n \left( v_{ix} - v_{nx} \right) \label{eq:neutral_momentumx},\\
& \frac{d}{dz} \left( \rho _n  v_{nz}^2 + P _n \right)= \alpha \rho _i \rho _n \left( v_{iz} - v_{nz} \right) \label{eq:neutral_momentumz},\\
& v_{nz}\frac{d P_n}{dz} + \gamma P_n \frac{d v_{nz}}{d z} = \left( \gamma -1 \right)\left( \Gamma - \Lambda \right) \label{eq:neutral_energy}
\end{align}
where $P_n$ is the neutral pressure, $\Gamma$ is the heating function, $\Lambda$ is the cooling function, and the internal energy $u = P_n / \left( \gamma - 1 \right)$ for adiabatic index $\gamma$. Finally, the electromagnetic equations give
\begin{align}
& \frac{d}{dz} \left( v_{iz} B_x - v_{ix} B_z \right) = 0 \label{eq:faradayx},\\
& \frac{d}{dz} \left( B_z \right) = 0. \label{eq:solenoid}
\end{align}
Solving for the internal structure of intermediate shocks requires equations analogous to Equations~\eqref{eq:ion_momentumx} to \eqref{eq:solenoid} for the $y$-direction. We consider only fast and slow shocks here by ignoring this case. Far away from the shock front there is no velocity difference between the fluids. This means there is no frictional heating in those regions, and so the shocks satisfy the isothermal one-fluid jump conditions. The types of shocks allowed are therefore determined by the MHD signal speeds.

\subsubsection*{Shock Families}

\begin{figure}
  \centering
  	\includegraphics[width=0.48\textwidth]{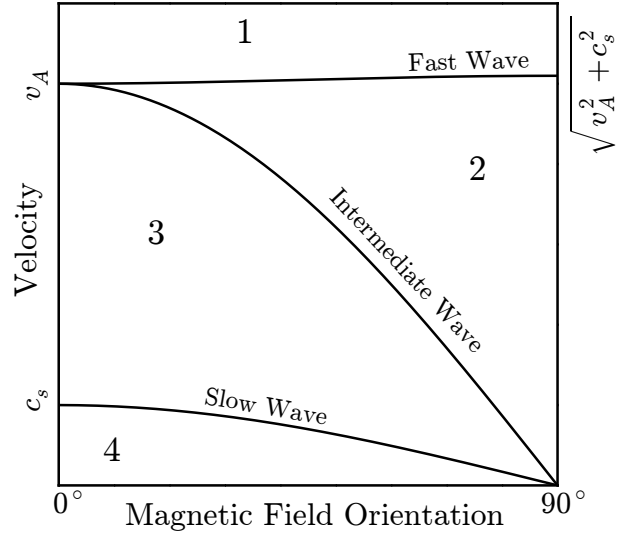}
 	\caption{The phase velocities of linear MHD wave modes versus the angle between the magnetic field and direction of propagation of those modes for $v_A > c_s$. The wave speeds delineate the regions marked $1$ to $4$.}
  \label{fig:ShockMap}
\end{figure}

When considering small perturbations of density, pressure, the velocity field, and the magnetic field around time independent averages the equations of MHD allow for three linear wave modes: the fast, intermediate and slow waves. These waves travel with phase velocities
\begin{align*}
f &= \left( \frac{v_A^2 + c_s^2}{2} + \frac{1}{2}\sqrt{\left( v_A^2 + c_s^2 \right)^2 - 4v_A^2c_s^2\cos ^2\theta} \right)^{1/2} ,\\
i &= v_A \cos \theta ,\\
s &= \left( \frac{v_A^2 + c_s^2}{2} - \frac{1}{2}\sqrt{\left( v_A^2 + c_s^2 \right)^2 - 4v_A^2c_s^2\cos ^2\theta} \right)^{1/2},
\end{align*}
where $v_A=B/\sqrt{4 \pi \rho}$ is the Alfv\'{e}n velocity, $c_s = \sqrt{k_B T/ \mu_m}$ is the isothermal sound speed with Boltzmann constant $k_B$ and mean mass per particle $\mu _m = \left( 7/3 \right) m_H$, and $\theta$ is the angle between the magnetic field and the direction of propagation under consideration. These speeds are plotted as functions of $\theta$ in Figure~\ref{fig:ShockMap} for the case $v_A > c_s$ (the relevant case in molecular clouds). 

In the frame of reference comoving with a shock front, the preshock fluid travels toward the shock front at a shock velocity $v_s$ greater than one of the wave speeds. The three wave speeds demarcate four regions of fluid velocities marked $1$ to $4$ in Figure~\ref{fig:ShockMap}. Inside of the shock front the fluid must transition down across a wave speed, which allows for six kinds of shock waves collected into three families: fast, intermediate and slow MHD shocks. Fast shocks are those that cross the fast wave speed only $\left( 1\rightarrow 2 \right)$, intermediate shocks cross the intermediate wave speed ($1\rightarrow 3$, $1\rightarrow 4$, $2\rightarrow 3$ and $2\rightarrow 4$), and slow shocks cross the slow wave speed only $\left( 3\rightarrow 4 \right)$. 

\begin{figure}
  \centering
    \includegraphics[width=0.48\textwidth]{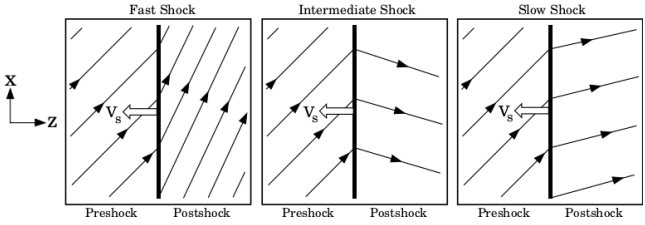}
    \caption{The effect on magnetic field orientation of the three classes of MHD shock waves. Fast shocks (left) increase the angle between the field and shock normal, intermediate shocks (middle) reverse the sign of the angle and slow shocks (right) decrease it.}
  \label{fig:ShockClasses}
\end{figure}

Far ahead and far behind a shock where there is no friction between the fluids---and hence $\textbf{v}_n = \textbf{v}_i$---one can show from Equations~\eqref{eq:ion_continuity} to \eqref{eq:solenoid} that the product
\begin{align*}
 \left(\frac{v_z^2 - i^2}{v_z}\right) B_x
\end{align*}
is conserved across any shock. This can be used to emphasise some basic differences between the different kinds of shocks. Suppose, for simplicity that $B_x>0$ in the preshock medium. Recall that a fast shock crosses the fast wave speed only, so that the term in parentheses remains positive but is reduced across the shock. Thus $B_x$ must increase to compensate. In intermediate shocks, the velocity crosses the intermediate wave speed and so the term in parentheses switches sign across the shock. This implies that $B_x$ must switch sign also. Finally, in slow shocks, the velocity crosses the slow wave speed only, so that the term in parentheses is negative and becomes further negative in the postshock medium. This means that $B_x$ must decrease to compensate. These three effects on the magnetic field direction are shown schematically in Figure~\ref{fig:ShockClasses}. 

The switch in sign of $B_x$ in the intermediate shock is due to a rotation of the magnetic field within the shock front. As the field rotates out of the $x$-$z$ plane, intermediate shocks require equations analogous to Equations~\eqref{eq:ion_momentumx} to \eqref{eq:solenoid} for the $y$-direction. This case is ignored here because it is unclear whether steady state intermediate shocks are physically admissable \cite[e.g.,][]{wu_mhd_1987, falle_inadmissibility_2001}. Furthermore, the $1 \to 3$ intermediate shock will resemble a fast shock as it crosses the fast speed, the heating in a $2\to 4$ shock front will be dominated by a hydrodynamic jump as it crosses the sound speed---see Section~\ref{ch:Integration}---and thus resemble a slow shock, the $2\to 3$ shock will only weakly heat the gas as it resembles a rotational discontinuity, and finally the $1\to 4$ shock will resemble a fast shock followed by a weak $2\to 3$ shock followed by a slow shock, so our models should also roughly capture its structure \citep{kennel_mhd_1989}. 

\highlight{As the spacing of field lines is proportional to the magnetic field strength}, one can see in Figure~\ref{fig:ShockClasses} that the field strength increases across fast shocks and decreases across slow shocks. This means that some of the kinetic energy of a fast shock is converted into magnetic field energy. Hence, for slow shocks at the same velocity one expects there to be more energy available to heat the gas.

\subsubsection*{Governing Differential Equations}

With some manipulation, Equations~\eqref{eq:ion_continuity} to \eqref{eq:solenoid} reduce to three ordinary differential equations determining changes in the $x$ component of the magnetic field and the neutral fluid temperature:
\begin{align}
\frac{d B_x}{dz} &= \frac{v_s^2 \alpha \rho_{i0}B_0^2}{v_A^2 B_z} \left( \frac{v_{ix}-v_{nx}}{v_{nz}v_{iz}} \right), \label{eq:ODE1}\\
\frac{d T_n}{dz} &= \frac{ T_n \left(\gamma -1 \right)}{\rho _{n0}  v_s^3\left( v_{nz}^2- \gamma  \tau v_s^2\right) } \times  \label{eq:ODE3}\\
&\left(\left( \frac{v_{nz}^2}{\tau} -v_s^2 \right)\left( \Gamma-\Lambda \right) -  \alpha \rho _i \rho _n \left( v_{iz} - v_{nz} \right) v_{nz}v_s^2\right) \nonumber
\end{align}
with the velocities
\begin{align*}
v_{nx} &= \frac{ v_A^2 B_z }{ v_s B_0^2 }  \left( B_x - B_{x0} \right),\\
v_{ix} &= \frac{ v_{iz} B_x - v_s B_{x0} }{ B_z },\\
v_{nz} &= \frac{ v_s }{ 2 }\left( \beta \pm \sqrt{\beta ^2 - 4\tau}\right),\\
v_{iz} &= \frac{ v_s B_{x0} B_x + B_z \left( B_x v_{nx} + B_z v_{nz} \right) }{ B_0^2 },
\end{align*}
the preshock Alfv\'{e}n velocity
\begin{align*}
v_A &= \frac{ B_0 }{ \sqrt{ 4 \pi \rho _{n0}  } }
\end{align*}
and
\begin{align*}
\beta &= 1 + \frac{ k_B T_{n0} }{ \mu _m v_s^2 } + \frac{1}{2}\frac{v_A^2}{v_s^2} \frac{B_{x0}^2 - B_x^2}{B_0^2},\\
\tau &= \frac{k_B}{\mu _m v_s^2} T_n
\end{align*}
where a subscript $0$ denote preshock values.

Equations~\eqref{eq:ODE1} and \eqref{eq:ODE3} are not complete without specifying how energy leaves the gas via $\Lambda$ and enters via $\Gamma$. The collisions between the ion and neutral fluids generate frictional heating at a rate
\begin{align*}
\Gamma _F = \alpha \rho _n \rho _i \left( \textbf{v}_i - \textbf{v} _n  \right) ^2.
\end{align*}
per unit volume. In addition, we include cosmic-ray heating, which is only important at temperatures of 10--30 K. 

The shock heated gas cools by radiating away heat energy stored in the rotational and vibrational modes of its constituents. These modes are collisionally excited, so that the cooling function must be a function of the densities of the coolants, the density of the colliding particles and the temperature of the gas. The abundances of coolants can change due to chemical reactions, and so we present a simple model for the oxygen chemistry occuring in the shock heated gas in the next section before specifying how that gas radiates in Section~\ref{sec:Cooling}.

\subsection{Chemistry}\label{sec:Chemistry}

Radiative cooling depends on the the abundances of the coolants and hence chemical reactions influence the cooling of the gas. The dominant coolants in molecular clouds are CO, H$_2$ and H$_2$O \citep[e.g.,][]{neufeld_radiative_1993} and the abundance of the latter is determined in this study by the set of eight reactions listed in Table~\ref{tab:RateCoefficients}, adopted from \cite{wagner_oxygen_1987}. Previous models of the gas phase chemistry in interstellar environments---using more than $100$ reactions---were found to be dominated by a small set of reactions \citep[e.g.,][]{iglesias_nonequilibrium_1978}. The reactions of \citeauthor{wagner_oxygen_1987} are the subset of these dominant reactions that control the abundance of H$_2$O.

\begin{figure}
  \centering
    \includegraphics[width=0.47\textwidth]{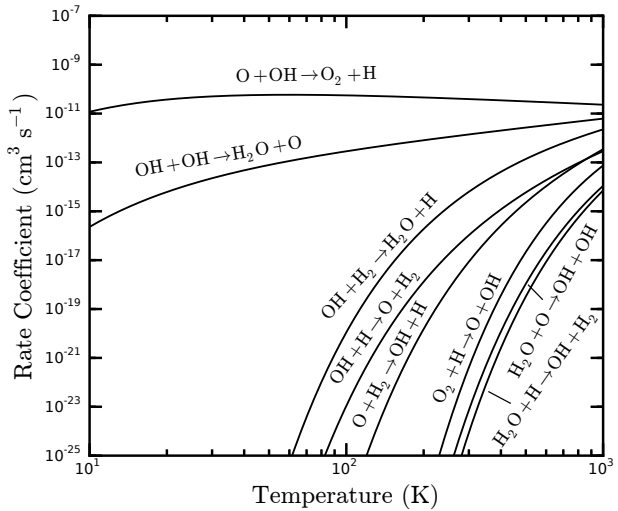}
    \caption{Reaction rate coefficients as a function of temperature.}
  \label{fig:ReactionRates}
\end{figure}

The change in the particle density of a neutral species $M$ through the shock is given by
\begin{align*}
\frac{d}{dz} \left( n(M)v_{nz} \right) = S_M
\end{align*}
where $S_M$ is the rate at which $M$ is created or destroyed. $\sum S_M$ over all neutral species must be zero in order for Equation~\eqref{eq:neutral_continuity} to remain true. The low densities of molecular clouds mean that only two body chemical reactions and photodissociation processes need to be considered. Two-body reaction rates between species $A$ with particle density $n(A)$ and species $B$ with particle density $n(B)$ take the form
\begin{align*}
k_{AB}(T) n(A)n(B) \text{ cm} ^{-3} \text{ s}^{-1}
\end{align*}
where $k_{AB}(T)$ is a rate coefficient. The UMIST Database for Astrochemistry 2012 (RATE12) \citep{mcelroy_umist_2013} gives this coefficient in the form
\begin{align*}
k_{AB}(T) = \alpha \left( \frac{ T }{ 300 } \right) ^\beta \exp \left( -\frac{ \gamma }{ T } \right) \text{ cm} ^3 \text{ s}^{-1}
\end{align*}
where $\alpha$, $\beta$ and $\gamma$ are constants. The values of these parameters for the reactions used are shown in Table~\ref{tab:RateCoefficients}. These rates are verified for the temperature range $200-2500$ K, so there is some uncertainty when extrapolating down to $10$ K. In fact, for the reaction
\begin{align*}
\text{O} + \text{OH}   \rightarrow \text{O}_2 + \text{H}
\end{align*}
the rate coefficient parameters in the RATE12 database give a rate that diverges in the $10$--$100$ K range. For this rate the parameters are taken from \cite{wagner_oxygen_1987}. In Figure~\ref{fig:ReactionRates}, the rates are plotted against temperature. It can be seen that most of the rates only ``turn on'' at temperatures $T \gtrsim 60$ K, which is why the weak fast shocks considered here only negligibly affect the molecular abundances.

\begin{table}
\caption{Reaction rate coefficient parameters.}
{\centering
\begin{tabular}{l l c c c } \hline\hline
\multicolumn{5}{c}{\vspace{-0.4cm}}\\ 
No. & Reaction & $\alpha$ & $\beta$ & $\gamma$\\ \hline
\multicolumn{5}{c}{\vspace{-0.4cm}}\\ 
$1$ & $\text{O} + \text{H}_2 			\rightarrow \text{OH} + \text{H} $ 			& $3.14\times 10^{-13}$ & $2.70$ & $3150.0$ \\
$2$ & $\text{OH} + \text{H}  			\rightarrow \text{O} + \text{H}_2$  		& $6.99\times 10^{-14}$ & $2.80$ & $1950.0$ \\
$3$ & $\text{OH} + \text{H}_2 		\rightarrow \text{H}_2\text{O} + \text{H}$ 	& $2.05\times 10^{-12}$ & $1.52$ & $1736.0$ \\
$4$ & $\text{H}_2\text{O} + \text{H} 	\rightarrow \text{OH} + \text{H}_2$ 		& $1.59\times 10^{-11}$ & $1.20$ & $9610.0$ \\
$5$ & $\text{OH} + \text{OH} 			\rightarrow \text{H}_2\text{O} + \text{O}$	& $1.65\times 10^{-12}$ & $1.14$ & $50.0$   \\
$6$ & $\text{H}_2\text{O} + \text{O} 	\rightarrow \text{OH} + \text{OH}$  		& $1.85\times 10^{-11}$ & $0.95$ & $8571.0$ \\
$7$ & $\text{O} + \text{OH}   		\rightarrow \text{O}_2 + \text{H}$  		& $4.33\times 10^{-11}$ & $-0.5$ & $30.0$   \\
$8$ & $\text{O}_2 + \text{H}  		\rightarrow \text{O} + \text{OH}$   		& $2.61\times 10^{-10}$ & $0$    & $8156.0$ \\ \hline
\end{tabular}
}
\label{tab:RateCoefficients}

\emph{Notes.} The parameters for reaction 7 are taken from \cite{wagner_oxygen_1987} instead of RATE12.
\end{table}

Photodissociation in molecular clouds is caused by ultraviolet radiation generated by secondary electrons in cosmic ray ionization events \citep{prasad_uv_1983}. The rates of photodissociation events per volume of species $A$ with particle density $n(A)$ take the form
\begin{align*}
p_A \frac{ \zeta _{CR}}{ 1 - \omega } n(A)\text{ cm} ^{-3}\text{ s}^{-1}
\end{align*} 
where $\zeta _{CR}$ is the cosmic ray ionization rate, $\omega$ is the albedo of the dust grains found in molecular clouds and $p_A$ is an efficiency constant. The ionization rate $\zeta _{CR}$ is set to $10^{-17}$~s$^{-1}$, the albedo $w$ is set to $0.6$, and the values of $p_A$ for the reactions considered---taken from \cite{gredel_cosmic-ray-induced_1989}---are shown in Table~\ref{tab:PhotoReactions}.

With these two rates and using the static, planar assumptions the abundance of species $M$ relative to the hydrogen nuclei density $x_M = n(M)/n_H$ changes as
\begin{align*}
\frac{d}{d z}\left( x_M \right) = \frac{n_H^2}{n_0 v_s}\left(\sum  k_{AB}(T)x(A)x(B) + \sum  p_A \frac{ \zeta _{CR}}{ 1 - \omega } x(A) \right)
\end{align*}
where $n_0$ is the preshock H-nuclei density and the sums are taken over reactions that either produce or destroy $M$.

The end result of this chemistry is to add five more differential equations---coupled to each other through densities and to Equations~\eqref{eq:ODE1} and \eqref{eq:ODE3} through temperature---to follow the abundances of H, O, OH, O$_2$ and H$_2$O. We follow \cite{pon_molecular_2012} in using an initial H$_2$O abundance of $10^{-7}$, O abundance of $5.45 \times 10^{-4}$ and a C abundance of  $1.4 \times 10^{-4}$ which is assumed to be entirely locked up in CO. The CO abundance is assumed to be constant throughout the shock, because it has a dissociation temperature higher than any temperature reached in these weak shocks. The initial H, OH and O$_2$ abundances are set to $10^{-4}$, $10^{-12}$ and $10^{-10}$ respectively. H$_2$ abundance is then computed through the shock using $x_{\text{H}_2} = (1/2)(1-x_H)$.

\begin{table}
\caption{Photodissociation rate efficiency.}
\begin{center}
\begin{tabular}{ l c } \hline\hline
\multicolumn{2}{c}{\vspace{-0.4cm}}\\ 
Reaction & $p_A$ \\ \hline
\multicolumn{2}{c}{\vspace{-0.4cm}}\\ 
$\text{H}_2\text{O} + \text{Photon} \rightarrow \text{OH} + \text{H}$ & $971$ \\
$\text{OH} + \text{Photon}   		\rightarrow \text{O} + \text{H} $ & $509$ \\
$\text{O}_2 + \text{Photon}  		\rightarrow \text{O} + \text{O} $ & $751$ \\ \hline
\end{tabular}
\end{center}
\label{tab:PhotoReactions}

\emph{Notes.} Photodissociation rate efficiencies taken from \cite{gredel_cosmic-ray-induced_1989}.
\end{table}

\subsection{Cooling}\label{sec:Cooling}

The cooling function of \cite{neufeld_radiative_1993} is adopted for this model. This uses an escape probability to account for the effects of reabsorption by the surrounding media on the rotational level populations. Collisions with H$_2$ are the only excitations considered as the abundance of H$_2$ in molecular clouds is orders of magnitude above the next most abundant molecular species. The power radiated by CO, H$_2$O and H$_2$ in a wide range of conditions is expressed in terms of a rate coeffecient $L_\text{M}$ defined such that the power radiated per unit volume by species M is
\begin{align*}
\Lambda _\text{M} = n(\text{M})n(\text{H}_2)L_\text{M}.
\end{align*}
$L_\text{M}$ is a function of H$_2$ density, temperature and an optical depth parameter $\tilde{N}(\text{M})$. $\tilde{N}(\text{M})$ is a correction factor that accounts for reabsorption of radiation by the same molecule in the surrounding gas. It depends on the geometry of the system in question, the density of M and the local velocity gradient. The expression for a plane-parallel slab of thickness $d$ and characteristic velocity difference $\Delta v$,
\begin{align}\label{eq:tildeN}
\tilde{N}(\text{M}) = \frac{ n(\text{M}) d }{9 \Delta v},
\end{align}
is used here. $L_\text{M}$ is then expressed with a four parameter analytic fit to its density dependence
\begin{align*}
\frac{1}{L_\text{M}} = \frac{1}{L_0} + \frac{ n(\text{H}_2) }{ L_{LTE} } + \frac{1}{L_0}\left( \frac{ n(\text{H}_2) }{ n_{1/2} }  \right) ^\alpha \left( 1 -  \frac{ n_{1/2} L_0 }{ L_{LTE} } \right)
\end{align*}
where $L_0$ is the low density limit of the cooling rate coefficient, $L_{LTE}$ is the luminosity per molecule with the level population in local thermodynamic equilibrium and $n_{1/2}$ is the H$_2$ density at which the cooling rate coefficient is half of $L_0$. $L_0$ is a function of temperature only while $L_{LTE}$, $n_{1/2}$ and $\alpha$ are functions of temperature and $\tilde{N}(\text{M})$.

Preliminary isothermal shocks were computed in order to gain an estimate on the values of $d$ and $\Delta v$, and these values checked for self consistency against the results when cooling was included. The average value of $n(M)$ through the shock was used. H$_2$ is optically thin in all interesting astrophysical conditions, so $L_{H_2}$ does not depend on Equation~\eqref{eq:tildeN}.

The values of $L_0$, $L_{LTE}$, $n_{1/2}$ and $\alpha$ for CO, H$_2$O and H$_2$ are taken from  \cite{neufeld_thermal_1995} in the temperature range $T=10$--$100$ K, and from \cite{neufeld_radiative_1993} in the range $T=100$--$2000$ K. We combine the parameters in the $T=10$--$100$ K range using an ortho-to-para H$_2$O ratio of 3:1, which was already assumed in the $T=100$--$2000$ K range. The values of some of these parameters at $T = 100$ K are inconsistent between the two sources, so we take the mean value.

\section{Numerical Integration}\label{ch:Integration}

The core of this problem are the first order ordinary differential equations which take the form:
\begin{align*}
\frac{d B_x}{dz} 	&= f_1\left( z, B_x, T \right)\\
\frac{d T}{dz} 		&= f_2\left( z, B_x, T \right)
\end{align*}
for $f_1$ and $f_2$ defined by Equations~(\ref{eq:ODE1}) and (\ref{eq:ODE3}). They must be solved simultaneously with the abundance derivatives which are coupled to these two through their dependence on temperature. When supplied with initial conditions, these equations can be integrated to give $B_x$, $T$ and the abundances as functions of $z$. The open source \emph{PYTHON} module \emph{scikits.odes}\footnote{https://github.com/bmcage/odes} was used, which solves initial value problems for ODEs using variable-order, variable-step, multistep methods.

The initial conditions are stationary states---in that their $z$ derivatives are zero---so integrating from these points changes nothing. A perturbation must be added to these initial conditions in the form
\begin{align*}
B_x &= B_{x0} + \delta B_x e^{\lambda _x z}\\
T &= T_0 + \delta T e^{\lambda _T z}
\end{align*}
before integrating. The effect of the perturbation can be understood by linearising the differential equations (\ref{eq:ODE1}) and (\ref{eq:ODE3}). It is illuminating to look at the isothermal case---where $\gamma = 1$---so that only the $B$ field derivative remains. In this case we get the eigenvalue
\begin{align*}
\lambda _x &=  \frac{\alpha  \rho _{i0}}{v_s}  \frac{\left(v_s^2 - f^2\right)\left(v_s^2 - s^2\right)}{v_A^2\left( v_s^2 - c_s^2\right)}
\end{align*}
where $f$, $i$ and $s$ are the fast, intermediate and slow signal speeds defined in Section~\ref{sec:MHD}. $\lambda _x$ determines whether the perturbation of $B_x$ grows or decays. By replacing the preshock variables with their postshock counterparts, the eigenvalue can also be used to explore the region of solution space near postshock states.

For fast shocks, the preshock state is in region $1$ of Figure~\ref{fig:ShockMap} so that $v_s > f > i > c_s > s$. This means $\lambda _x$ is positive so that $B_x$ grows. This describes an unstable stationary point, where any perturbation away from the initial condition grows. The postshock state is in region $2$ so that $f > v_f > i > c_s > s$ where $v_f$ is the final velocity. This means $\lambda _x$ is negative so that this state is a stable stationary point. In Section~\ref{sec:MHD}, we noted that the fast shock increases $B_x$, hence a positive perturbation of the initial $B_x$ is all that is required to finish at the fast postshock state.

For slow shocks, the preshock state is in region $3$ of Figure~\ref{fig:ShockMap} so that $f > i > v_s > c_s > s$ in the supersonic case. This means $\lambda _x$ is negative and this state is a stable stationary point. Hence there is no way to leave the supersonic slow preshock state in a continuous fashion like in C-type fast shocks. The slow solution is further complicated by a singularity in the equations when crossing the sound speed $\left(v_{nz} \to c_s \right)$, as a manipulation of Equation~(\ref{eq:neutral_energy}) gives
\begin{align*}
\frac{d v_{nz}}{dz} = \frac{\left( \gamma -1 \right) \left( \Gamma - \Lambda \right) - \alpha \rho _n \rho _i \left( v_{iz} - v_{nz} \right) v_{nz}}{\rho _n \left( c_s^2 - v_{nz}^2 \right) }.
\end{align*}
We cross this sonic point by inserting a gas dynamic jump in the neutrals determined by the hydrodynamic jump conditions:
\begin{align}
\frac{v_2}{v_1} &= \frac{\gamma -1}{\gamma +1} + \frac{2}{\gamma + 1} \frac{1}{M^2}  \label{eq:VelJump}\\
\frac{T_2}{T_1} &= \left( 1 + \frac{2 \gamma}{\gamma + 1}\left( M^2 -1 \right) \right) \frac{M^2 \left( \gamma -1 \right) +2}{M^2\left( \gamma +1 \right)} \label{eq:TempJump}
\end{align}
where $M^2 = \rho _1 v_1^2 / \gamma P_1$ and $v_2$ and $T_2$ are the neutral $z$ velocity and temperature immediately after the gas dynamic jump. The slow postshock state lies in region $4$ of Figure~\ref{fig:ShockMap} so that $f > i > c_s > s > v_f$ and $\lambda _x$ is negative. This means this state is a stable stationary point, and so jumping across the sound speed (via Equation~(\ref{eq:VelJump})) will allow the solution to smoothly settle onto the slow postshock state.

For slow shocks with a subsonic preshock state, $f > i > c_s > v_s > s$ so that $\lambda _x$ is positive and the stationary state is an unstable. This means a perturbation that reduces $B_x$ will grow smoothly until the solution reaches the slow postshock state. Such a shock is a C-type slow MHD shock, and will be ignored here because it requires high sound speeds and therefore high temperatures that are not obviously relevant for molecular cloud studies.

%\begin{figure}
%\centering 	
%	\includegraphics[width=0.48\textwidth]{scripts/PeakTemp/PeakTemp.pdf}
%	\caption{Postshock temperature in slow shocks. This plot shows the temperature that a slow shock would reach given a shock velocity determined by \Cref{eq:TempJump} for preshock temperature of $10$ K and adiabatic index $\gamma = 5/3$.}\label{fig:PeakTemp}
%\end{figure}

\section{Results}\label{sec:Results}

\begin{figure}
\centering 	
	\includegraphics[width=0.48\textwidth]{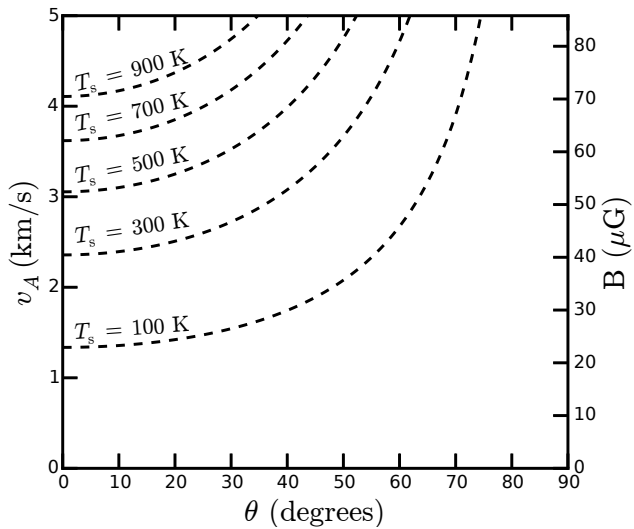}
	\caption{Alfv\'{e}n velocity versus magnetic field orientation, with dashed lines of constant intermediate speed $v_A \cos \theta$ that allow slow MHD shocks to reach the peak temperatures shown. The magnetic field strength on the second vertical axis assumes preshock density $n_0=10^3$~cm$^{-3}$.}
	\label{fig:ParameterSpace}
\end{figure}

\begin{table} 
\caption{Shock Parameters.} 
\centering
\begin{tabular}{l c c c} \hline\hline
Parameters & $v_s$ & $\log \tilde{\rm{N}}$(CO) & $\log \tilde{\rm{N}}$(H$_2$O) \\  
 & km/s & \multicolumn{2}{c}{$\log \left( \rm{cm}^{-2} \, \left( \rm{km/s} \right) ^{-1} \right)$} \\ \hline 
\multicolumn{4}{c}{Fast Shocks, $\theta = 89.9^\circ$} \\ \hline
$n_0 = 10^2$ cm$^{-3}$ & 2.0 & 15.9 & 11.7\\ 
$B_0 = 3 \mG$ & 2.5 & 15.7 & 11.8\\ 
 & 3.0 & 15.6 & 11.9\\
 & 3.5 & 15.5 & 11.9\\ 
 & 4.0 & 15.5 & 11.9\\ \hline
$n_0 = 10^3$ cm$^{-3}$ & 2.0 & 15.8 & 12.7\\ 
$B_0 = 10 \mG$ & 2.5 & 15.8 & 12.7\\ 
 & 3.0 & 15.8 & 12.6\\ 
 & 3.5 & 15.8 & 12.6\\ 
 & 4.0 & 15.5 & 12.3\\ \hline
$n_0 = 10^4$ cm$^{-3}$ & 2.0 & 15.9 & 12.8\\ 
$B_0 = 32 \mG$ & 2.5 & 15.8 & 12.7\\ 
 & 3.0 & 15.8 & 12.8\\ 
 & 3.5 & 15.7 & 12.6\\ 
 & 4.0 & 15.7 & 12.6\\ \hline
\multicolumn{4}{c}{Slow Shocks, $\theta = 30^\circ$} \\ \hline
$n_0 = 10^2$ cm$^{-3}$ & 2.0 & 14.5 & 11.0\\ 
$B_0 = 25 \mG$ & 2.5 & 14.5 & 11.0\\ 
 & 3.0 & 14.5 & 11.0\\ 
 & 3.5 & 14.5 & 11.4\\ 
 & 4.0 & 14.5 & 12.3\\ \hline
$n_0 = 10^3$ cm$^{-3}$ & 2.0 & 14.5 & 11.0\\ 
$B_0 = 80 \mG$ & 2.5 & 14.5 & 11.0\\ 
 & 3.0 & 14.5 & 11.5\\ 
 & 3.5 & 14.5 & 12.6\\ 
 & 4.0 & 14.5 & 13.2\\ \hline
$n_0 = 10^4$ cm$^{-3}$ & 2.0 & 15.6 & 12.6\\ 
$B_0 = 253 \mG$ & 2.5 & 15.7 & 12.7\\ 
 & 3.0 & 15.1 & 13.6\\ 
 & 3.5 & 15.0 & 14.6\\ 
 & 4.0 & 15.0 & 15.1\\ \hline 
\end{tabular}
\label{tab:ShockParams}
\end{table}

Here we compare a set of shock models with parameters shown in Table~\ref{tab:ShockParams}. The turbulent cascade of energy ensures the dissipation is dominated by low-velocity shocks, so we follow \cite{pon_molecular_2012} in looking at velocities around $3 \kps$. In Figure~\ref{fig:ParameterSpace}, the Alfv\'{e}n velocity is plotted against the magnetic field orientation $\theta$ with dashed lines of constant intermediate speed $i = v_A \cos \theta$ overlaid. The magnetic field strength on the second vertical axis is computed for a preshock density $n_0 = 10^3$ cm$^{-3}$. The intermediate speed is the upper bound of possible slow shock velocities---region 3 in Figure~\ref{fig:ShockMap}---and so the dashed lines in Figure~\ref{fig:ParameterSpace} trace the minimum field strength required to get slow shocks with peak temperatures shown. We thus chose $\theta$ low enough to retain reasonable field strength values. At $\theta = 30^\circ$, a $4 \kps$ slow shock---which will heat the gas to $\sim 857$ K immediately after the neutral jump---requires an Alfv\'{e}n velocity of $\sim 4.6 \kps$, which is fixed for all the slow shocks at different preshock densities. This gives magnetic field strengths in the range $B_0 = 25$--$253 \mG$. Finally, $\tilde{N}$ was initially chosen for preliminary isothermal shocks, and then recomputed after each shock model for consistency. Table~\ref{tab:ShockParams} shows the self consistent values of $\tilde{\rm{N}}$(CO) and $\tilde{\rm{N}}$(H$_2$O).

\subsubsection*{Structural characteristics}

\begin{figure}
\centering 	
	\includegraphics[width=0.5\textwidth]{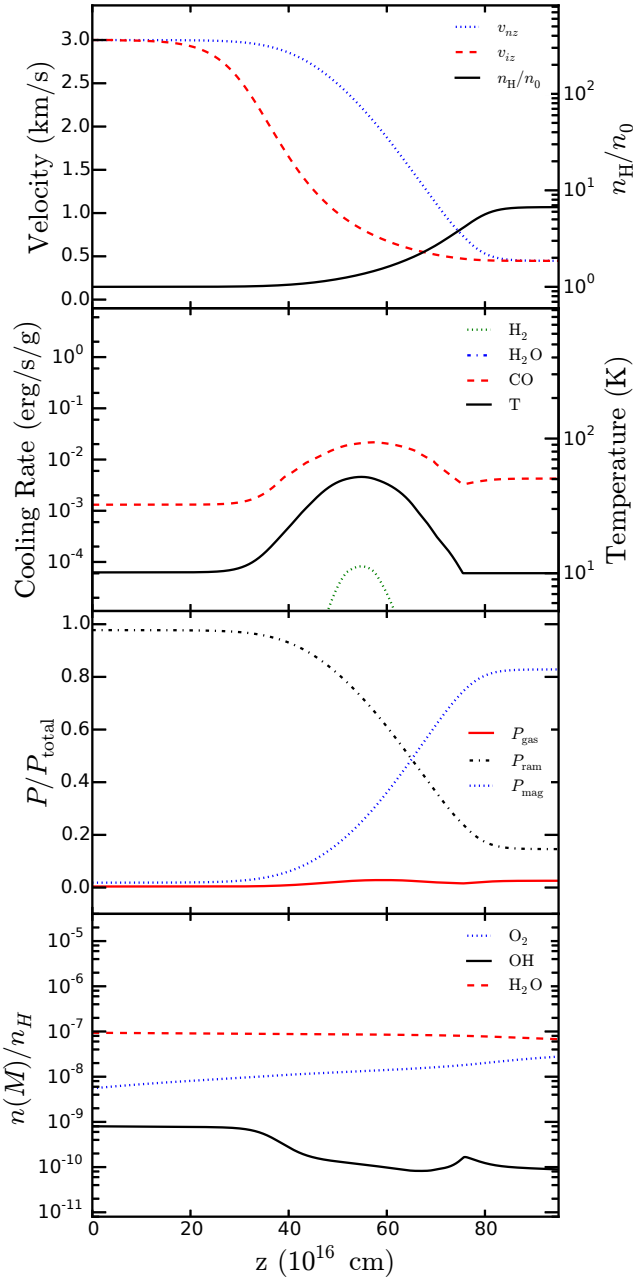}
	\caption{\highlight{Profiles for a $v_s = 3 \kps$ fast MHD shock propagating through a medium with initial number density $n_0 = 10^3$ cm$^{-3}$ at $89.9^\circ$ to a $10 \mG$ magnetic field. The first panel shows the velocity and density profiles, the second panel shows the temperature and cooling rate profiles, the third panel shows the gas, ram and magnetic pressure profiles, and the fourth panel shows the abundances of selected oxygen molecules.}}\label{fig:profile-fast}
\end{figure}

\begin{figure}
\centering 	
	\includegraphics[width=0.5\textwidth]{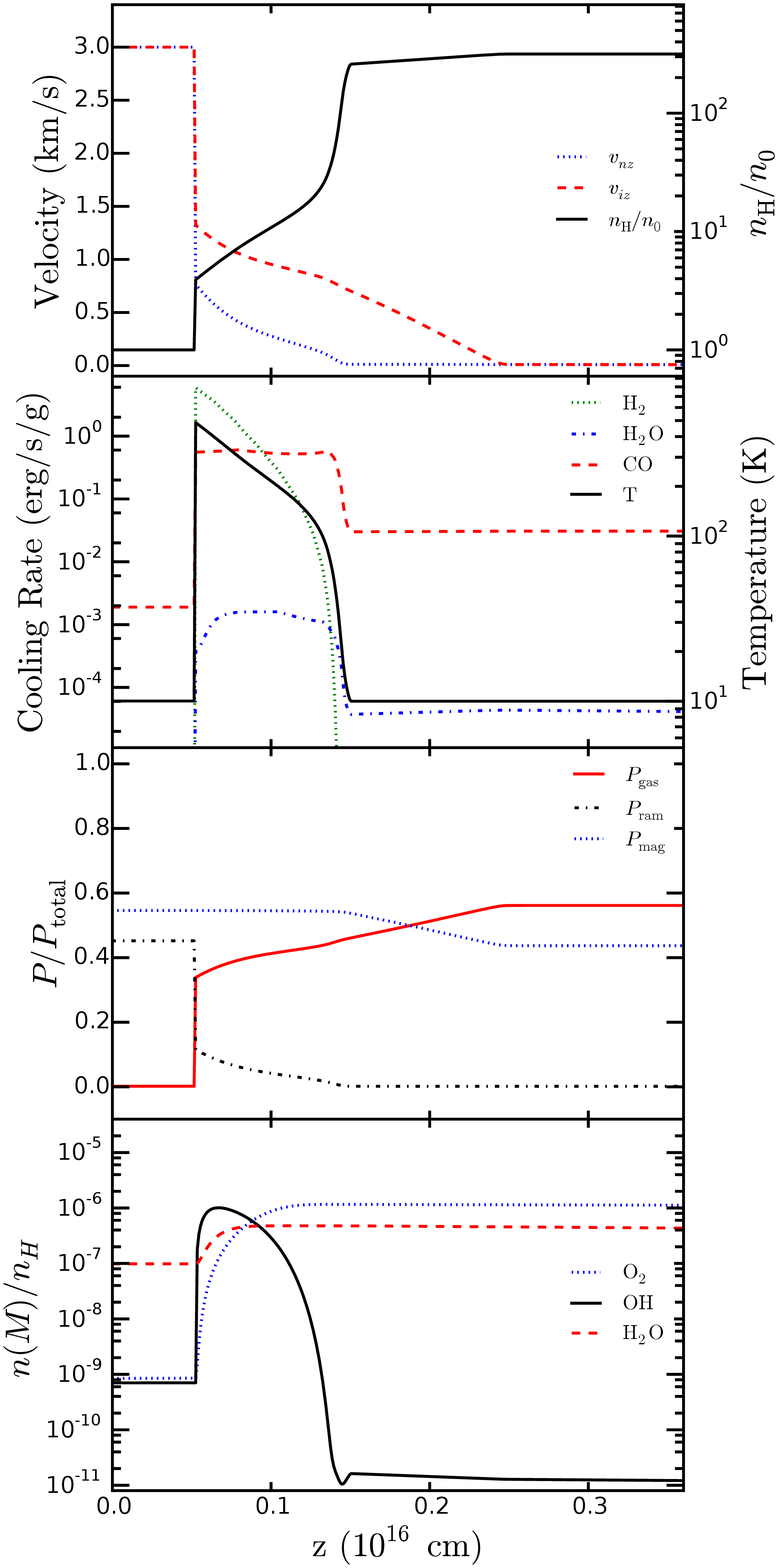}
	\caption{\highlight{Same as Figure~\ref{fig:profile-fast} but for a $v_s = 3 \kps$ slow shock propagating through a medium with initial number density $n_0 = 10^3$ cm$^{-3}$ at $30^\circ$ to an $80 \mG$ magnetic field.}}\label{fig:profile-slow}
\end{figure}

In Figures~\ref{fig:profile-fast} and \ref{fig:profile-slow} we compare fast and slow shock profiles of ion and neutral velocity, density, temperature, cooling rate and abundances for $v_s = 3\kps$. The fast shock propagates at $89.9^\circ$ to a $10 \mG$ magnetic field, while the slow shock propagates at $30^\circ$ to an $80 \mG$ field. The preshock density $n_0 = 10^3$ cm$^{-3}$ for both.

In the velocity profiles---\highlight{upper panels of Figures~\ref{fig:profile-fast} and \ref{fig:profile-slow}}---the neutral and ion velocities in the shock propagation direction are the dashed and dotted lines respectively. These plots are in the frame of reference of the shock wave, and so velocities below the shock velocity represent fluid flowing ahead of the shock front in the lab frame, in which the shock wave travels to the left. Therefore, in the fast shock the ions stream ahead, imparting some of their momentum to the neutrals until both fluids are moving at the same speed in the postshock medium (on the right side of each plot). This process is set by the long ion-neutral collision timescale and therefore happens smoothly over a large distance, and there is no viscous jump as seen in hydrodynamic shocks. In contrast, for the slow shock it is the neutrals that flow ahead of the ions that then get accelerated to the neutral velocity over a small distance---set by the fast neutral-ion collision timescale---giving slow shocks a much thinner structure than the fast shocks. The cooling timescale of the gas lies between the two collision timescales, and so slow shocks will reach higher peak temperatures than fast shocks at the same velocity.

The density profiles---solid black lines in the upper panels---similarly show large differences between the shocks. The fast shock weakly compresses the gas by a factor of $\sim 7$ in a simple, smooth manner. The slow shock has a complex density profile that ends with a compression ratio of $\sim 300$ and must be understood in conjunction with the temperature and cooling (see middle panel) taking place. It starts with the neutral jump, \highlight{governed by Equations~(\ref{eq:VelJump}) and (\ref{eq:TempJump})}, compressing the gas by a factor of $\sim 4$ and heating it to $\sim 500$ K. Combining Equations~(\ref{eq:ion_pressureB}) and (\ref{eq:neutral_momentumz}) we get
\begin{align*}
\frac{d}{dz}\left( \rho _n v_{nz}^2 + P_n + \frac{B^2}{8\pi} \right) = 0
\end{align*}
where we identify each term as effective pressure terms: ram pressure $P_{ram} = \rho _n v_{nz}^2$, gas pressure $P_{gas} = P_n = n k_B T$ and magnetic pressure $P_{mag} = B^2 / 8\pi$. As the neutrals lead the ions in the slow shock, Equation~(\ref{eq:ion_pressureB}) implies that the magnetic pressure must drop. The velocity drop causes the ram pressure to drop, and so gas pressure must increase, which is shown in the third panel of Figure~\ref{fig:profile-slow}. Efficient line emission from CO and H$_2$ causes the temperature to quickly decrease, so to compensate the density must increase. This occurs smoothly in the $\sim 0.1 \times 10^{16}~\rm{cm}$ after the initial jump, until the density reaches a plateau at a compression ratio of $\sim 300$. The density finally settles as the neutral and ion velocities equalise.

While the velocity and density structural differences are large, the observational implications are sensitive to the temperature and cooling rate profiles shown in the \highlight{second panels of Figures~\ref{fig:profile-fast} and \ref{fig:profile-slow}}. These panels show that the peak temperatures differ by an order of magnitude with the slow shock peaking at $\sim 487$ K and the fast shock peaking at $\sim 50$ K. At this temperature, the fast shock causes CO emission only slightly above the background level. The slow shock is hot enough to emit significantly in lines of CO and H$_2$, and also \highlight{in H$_2$O above the CO background emission. The fourth panels of Figures~\ref{fig:profile-fast} and \ref{fig:profile-slow} also show that the slow shock drives more chemistry than the fast shock due to its high temperature. The strong H$_2$O emission in the slow shock is partially due to this increase in its abundance.}

Following the pressure terms---in the \highlight{third panels of Figures~\ref{fig:profile-fast} and \ref{fig:profile-slow}}---through the shocks provides an intuitive understanding of the fundamental differences between fast and slow shocks. In the fast shock, the magnetic pressure (dotted lines) is higher in the postshock region than in the preshock region. As the gas pressure (solid lines) is neglible everywhere in this shock it is understood to be magnetically driven, where the high $P_{mag}$ region pushes into the low $P_{mag}$ region. The ion fluid is strongly coupled to the magnetic field, and so the ions are pushed forward ahead of the neutrals. This is why the ions lead the neutrals in the velocity profile. This situation is reversed in the slow shock, where a high gas pressure region pushes the neutrals through the ions. In this case, the ion coupling to the magnetic field deforms it and increases the separation between field lines, which reduces the field strength.

\subsubsection*{Comparison of fluxes}

\begin{figure}
\centering 	
	\includegraphics[width=0.48\textwidth]{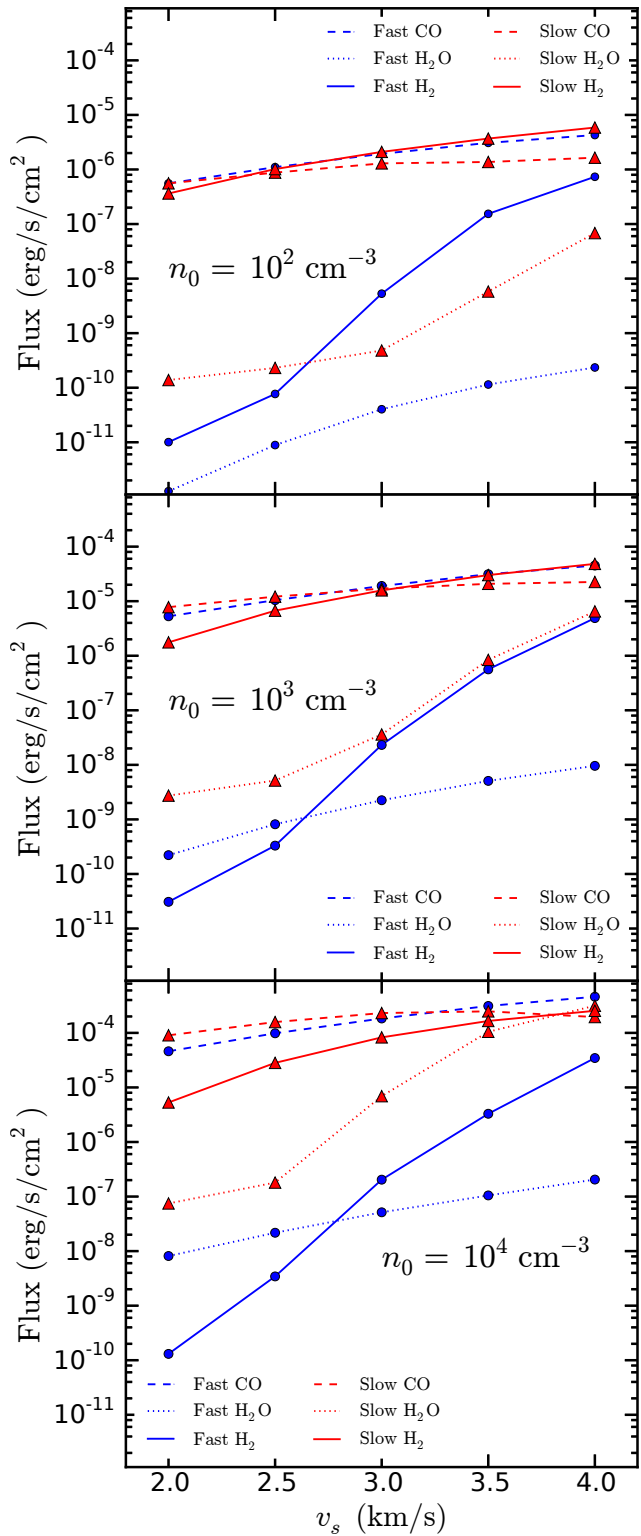}
	\caption{Energy flux contributions by different coolants in shocks of initial density $n_0 = 10^2$~cm$^{-3}$ (top), $n_0 = 10^3$~cm$^{-3}$ (middle) and $n_0 = 10^4$~cm$^{-3}$ (bottom), and shock velocities $v_s = 2$--$4 \kps$. Blue circles refer to fast shocks and red triangles refer to slow shocks. Solid, dotted and dashed line types indicate whether the coolant is CO, H$_2$ or H$_2$O respectively.}\label{fig:flux}
\end{figure}

The cooling discussed in Section~\ref{sec:Cooling} can be used to search for observational differences between the kinds of shocks. Integrating the cooling rate---middle panels of Figures~\ref{fig:profile-fast} and \ref{fig:profile-slow}---through the shocks gives the flux emitted in the direction of the shock normal. In Figure~\ref{fig:flux}, we compare the CO, H$_2$ and H$_2$O fluxes of all the fast and slow shocks in Table~\ref{tab:ShockParams}. 

The CO flux is similar between both kinds and across the velocity range. The H$_2$ flux is much more temperature sensitive and is therefore stronger in the slow shocks by 5 orders of magnitude at the lowest velocity and 1 order of magnitude at the highest. The H$_2$O flux is also stronger in the slow shock at all velocities.

Figure~\ref{fig:flux} shows that for slow shocks the CO and H$_2$ fluxes are within an order of magnitude of each other at all velocities. For fast shocks, however, the CO flux is always stronger by more than an order of magnitude. This feature holds at initial densities $n_0 = 10^2$ and $10^4$ cm$^{-3}$, though the magnitudes of the fluxes are lower and higher respectively by factors of $~10$ than the $n_0 = 10^3$ cm$^{-3}$ case. This suggests that combining observations of flux from both molecules is a strong indicator of the kind of shock being observed.

\subsubsection*{Rotational lines}\label{sec:radex_lines}

\begin{figure*}
\centering   	
	\includegraphics[width=\textwidth]{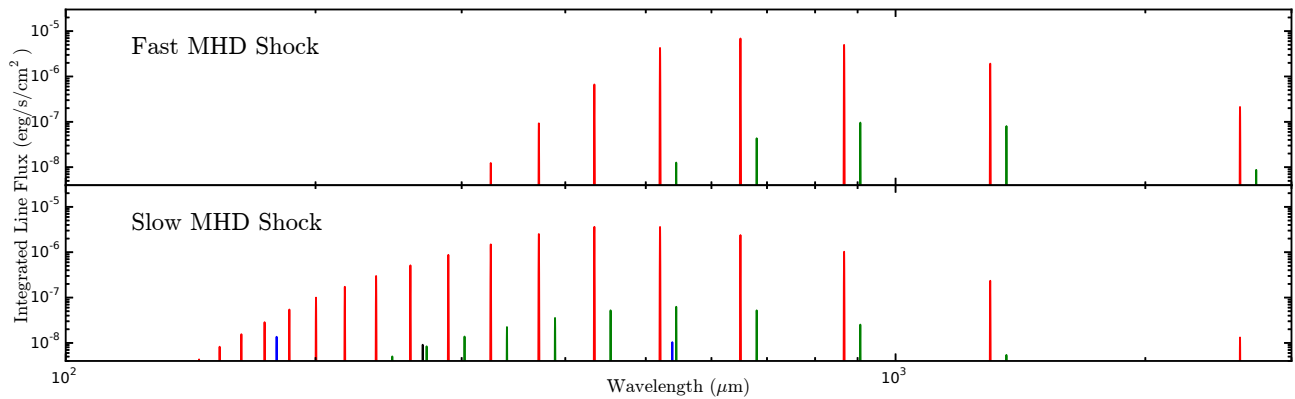}
	\caption{Estimates of integrated line fluxes from rotational lines of $^{12}$CO (red), $^{13}$CO (green), ortho-H$_2$O (blue) and para-H$_2$O (black) for the fast (top) and slow (bottom) shocks shown in Figures~\ref{fig:profile-fast} and \ref{fig:profile-slow}.}\label{fig:lines}
\end{figure*}

\begin{figure}
\centering   	
	\includegraphics[width=0.48\textwidth]{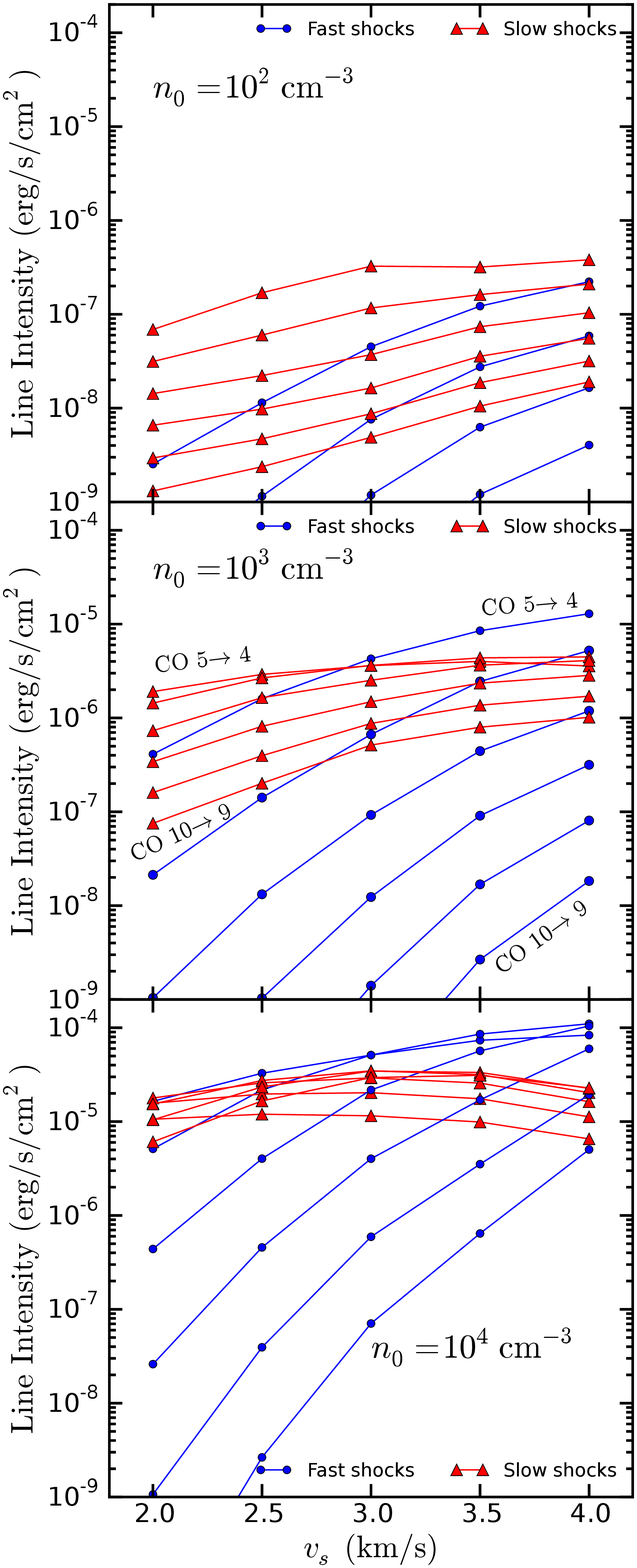}
	\caption{Integrated line intensities of CO rotational lines for fast (blue circles) and slow (red triangles) shocks with preshock densities $n_0 = 10^2 \, \rm{cm}^{-3}$ (top), $n_0 = 10^3 \, \rm{cm}^{-3}$ (middle) and $n_0 = 10^4 \, \rm{cm}^{-3}$ (bottom), and shock velocities $v_s = 2-4 \kps$. The solid lines connect rotational lines of the same transition from $J = 5\to 4$ to $J = 10\to 9$.}
	\label{fig:intensities}
\end{figure}

Figure~\ref{fig:flux} shows similar levels of CO flux for both fast and slow shocks at all the velocities considered here. However, the strength of the individual rotational lines that make up this flux is not expected to be the same given the higher temperatures that slow shocks reach. Here we estimate those those line strengths using the non local thermodynamic equilibrium (LTE) radiative transfer code RADEX \citep{van_der_tak_computer_2007}.

For a given radiating molecule, RADEX requires as input the density of H$_2$ as the collisional partner, the column density of the radiating molecule, the temperature and the linewidth. We consider 4 molecules with H$_2$ as the collisional partner: $^{12}$CO, $^{13}$CO, ortho-H$_2$O and para-H$_2$O. We use a $^{12}$CO to $^{13}$CO ratio of 61:1, well within the range of measurements given by \cite{milam_12c/13c_2005}, and an ortho-to-para H$_2$O ratio of 3:1 as assumed in Section~\ref{sec:Cooling} for the cooling function. 

We use RADEX in slab mode, and use density weighted averages of the required inputs over appropriate slab definitions of the computed shocks. For slow shocks, we define multiple slabs in order to account for the complex temperature and density structure. While RADEX accounts for optical depth effects within each slab and outputs the optical depth for each rotational transition, the emission from a slab may be reabsorbed by the other slabs it has to pass to reach the observer. Hence estimates of line strengths from lines of high optical depth---such as the $\rm{CO}$ lines below $J=5\to 4$---may not be reliable. \highlight{Differences in the optical depth of particular lines in fast and slow shocks may have an observational effect. For example, \cite{burkhart_turbulence_2013} show that the slope of the spatial power spectrum derived from synthetic observations of $^{13}$CO~$J=2\to 1$ in simulations of MHD turbulence is sensitive to the optical depth of this line. Furthermore, this slope could distinguish between sub- and super-Alfv\`{e}nic turbulence. However, to relate the optical depth of a line emitting from dense postshock regions to that over a line of sight through a cloud requires knowledge of the spatial distribution of shock waves in the cloud, which is beyond the scope of this paper. Finally, we normalise the line strengths so that emission from each molecule from all its lines is equal to the fluxes given in Figure~\ref{fig:flux}.}

In Figure~\ref{fig:lines} we plot the line strengths as computed by RADEX for the fast and slow shocks shown in Figures~\ref{fig:profile-fast} and \ref{fig:profile-slow}. As expected, the higher temperature of the slow shock results in strong excitation of the high-$J$ CO transitions of both isotopes. There are also excited H$_2$O lines in the slow shock that are negligible in the fast shock.

Any line of sight through a cloud will intersect multiple shocks propagating at different velocities, so we look for features in the spectra that hold across the velocity range. In Figure~\ref{fig:intensities} we plot the line integrated fluxes for selected high-$J$ lines of $\rm{CO}$ across the velocity and density ranges for fast and slow shocks. In all slow shocks, the dynamic range of these high-$J$ lines of $\rm{CO}$ is much lower than in fast shocks. For example, the line ratio CO~$J= 5 \to 4/10 \to 9$ for a fast shock is always greater than 30 times the ratio from the slow shock of the same velocity and density.

As line ratios don't change if the line strengths are reduced by a constant factor, these distinguishing features will remain even though the shocks propagate in various directions with respect to the line of sight. Furthermore, any line of sight through a turbulent cloud will cross multiple shock fronts. Emission from optically thin lines will be a simple addition of emission from each shock, retaining line ratio characteristics which therefore could be strong indicators of shock type.

\highlight{The high temperatures reached within slow shocks allow H$_2$ cooling to become comparable to cooling by CO, as can be seen in Figure~\ref{fig:flux}. The H$_2$ molecule has no dipole moment, and so this cooling is due to weak quadrupole emission. We can estimate the low lying pure rotational lines ($\nu = 0 \to 0$) of this emission using Figure 1 of \cite{burton_mid-infrared_1992}, which assumes the column of radiating H$_2$ is in LTE. For pure rotational lines S(0)-S(3), the line integrated fluxes emitted normal to the plane of the slow shock in Figure~\ref{fig:profile-slow} are in the range $\sim 10^{-7}$-$10^{-6}$ erg/cm$^2$/s. These lines a strongly suppressed in gas below 100 K, and so the fast shock in Figure~\ref{fig:profile-fast} produces negligible S(0)-S(3) line emission. This estimate shows that the pure rotational lines of H$_2$ could be an important diagnostic of shock type. However, the assumption of LTE doesn't always hold and so to produce a more accurate prediction for H$_2$ rovibrational line emission the level populations would have to be computed in parallel with the shock flow variables and reaction rate equations, as is done for instance in the fast shock model of \cite{gusdorf_sio_2008}.}
\section{Discussion}\label{sec:Discussion}

\highlight{We have integrated the two-fluid MHD equations to obtain one dimensional time-independent fast and slow shock wave solutions. While two-fluid fast MHD shocks have been well studied \citep[e.g.][]{draine_multicomponent_1986,flower_c-type_1998}, two-fluid slow shocks have not been considered in molecular cloud conditions. Thus we use a simplified model in order to highlight the qualitative differences between the two kinds of shocks. Simplifications include restricting the chemical network to a small subset of reactions that influence the abundance of H$_2$O, as this is an important coolant in molecular gas. Gas-phase oxygen chemistry is well contained in the restricted network used here \citep{iglesias_nonequilibrium_1978} and so a richer network---such as that used in \citep{glover_modelling_2010}---would not strongly affect H$_2$O production. We have not modeled how a variety of different initial abundances---which varies throughout a turbulent molecular cloud \citep{glover_modelling_2010}---could affect the shock chemistry or cooling profiles. However, this will not change the general differences in structure between fast and slow shocks which is determined by whether the shock is driven by magnetic or gas pressure, respectively. Rotational line emission from H$_2$ could provide more observational diagnostics and has been computed in fast shock models by \cite{lesaffre_low-velocity_2013}. In order to predict this emission accurately the level populations of the rovibrational states of H$_2$ have to be computed in parallel with the shock flow variables and reaction rate equations.}

\highlight{This work shows that fast and slow MHD shocks are structurally and observationally distinct. In Section~\ref{sec:MHD} we noted that fast shocks increase the angle between the magnetic field and direction of propagation whereas slow shocks decrease this angle. In principle this effect could be observed in studies of the polarized thermal emission revealing the geometry of magnetic fields in molecular clouds. The study by \cite{planck_collaboration_planck_2015} has shown that in high column density filaments the magnetic field tends to be perpendicular to the filament. \cite{padoan_turbulent_2001} explain these filaments as dense postshock regions resulting from the collision of supersonic turbulent flows. This result precludes the possibility that the filaments contain the fast shocks most commonly modeled \citep[e.g.][]{flower_excitation_2010, pon_molecular_2012} in which the magnetic field is parallel to the shock front. Higher spatial resolution work of this kind could either detect the magnetic field bending across shock fronts or rule out the shock formation scenario of filaments. In Section~\ref{sec:radex_lines} we showed that observations of high-$J$ CO lines---above $J=6 \to 5$---could distinguish between fast and slow MHD shocks. Multiple transitions should simultaneously be observed to disentangle the possible shock models through the use of line ratios. In addition, velocity information from linewidths could be used to constrain the shock velocity of the model. Hence, useful observations would require spectral resolutions better than 1 km/s at the high-$J$ CO lines (CO $J=6 \to 5$ to $J=10 \to 9$ lines lie in the frequency range 691.47--1151.99 GHz). In addition, the spatial resolution must be sufficient to avoid gas warmed by protostellar outflows or stellar winds. By way of example, we consider a recent observation that satisfies these criteria and show that slow shock signatures may be present.}

\citet{pon_mid-j_2015} (hereafter P15) present observations of CO~$J= 8 \to 7$, $9 \to 8$ and $10 \to 9$ taken with the \textit{Herschel} Space Observatory, towards four starless clumps within Galactic infrared dark clouds (IRDCs). These clumps were chosen because they lacked massive embedded protostars, avoiding confusion from outflows which could also create a warm gas component. The authors detect CO~$J= 8 \to 7$ and $9 \to 8$ towards three of their clumps---named C1, F1 and F2---and give an upper limit for the $10 \to 9$ line. They compare these observations to PDR models at densities of $10^4$ and $10^5 \, \rm{cm}^{-3}$, typical of IRDCs, and interstellar radiation fields of 1 and $\sim 3$ Habing. All of the PDR models underpredict the CO~$J= 9 \to 8$ line, so the authors suggest that the dissipation of turbulence in low velocity shocks could account for these lines. Here we consider how fast and slow shocks similar to those modeled in this paper could account for these observations.

\begin{table*}
 \caption{CO high-$J$ line luminosities for IRDCs \citep{pon_mid-j_2015} and selected shock models}
 \label{tab:IRDClines}
 \begin{tabular}{c c c c} \hline\hline
 \multicolumn{4}{c}{\vspace{-0.4cm}}\\
 Source & CO $J= 8 \to 7$ & CO $J= 9 \to 8 / 8 \to 7$ & CO $J= 10 \to 9 / 8 \to 7$ \\
  & $\left(10^{-14} \, \rm{erg/s/cm}^2 \right)$ & & \\ \hline
 \multicolumn{4}{c}{\vspace{-0.4cm}}\\
 C1 & $1.26 \pm 0.06$ & $0.65 \pm 0.07$ & $< 2.10 \pm 0.24$\\
 F1 & $1.48 \pm 0.09$ & $0.94 \pm 0.10$ & $< 2.20 \pm 0.29$\\
 F2 & $1.19 \pm 0.09$ & $0.77 \pm 0.10$ & $< 2.69 \pm 0.36$ \\
 Slow Shock Models & & & \\
 (A) & 1.66 & 0.51 & 0.27 \\
 (B) & 1.75 & 0.80 & 0.58 \\
 Fast Shock Models & & & \\
 (A) & 1.19 & 0.16 & 0.03 \\
 (B) & 4.10 & 0.25 & 0.06 \\ 
 \hline
 \end{tabular}\\
 \emph{Notes.} The values used for the $J= 10 \to 9$ line from IRDCs are the suggested upper limits. Slow shocks A and B have shock velocities $v_s = 2 \kps$ and $3.5 \kps$ respectively. Fast shocks A and B have shock velocities $v_s = 3.5 \kps$ and $4.0 \kps$ respectively. All four shock models have preshock density $n_0 = 10^4 \, \rm{cm}^{-3}$.
\end{table*}

To convert the integrated line intensity ($\int T dv$) of P15 to flux we use the formula
\begin{align*}
F = \frac{2 k \Omega \nu^3}{c^3} \int T dv
\end{align*}
where $k$ is the Boltzmann constant, $\Omega$ is the beam area, $\nu$ is the frequency of the transition under consideration and $c$ is the speed of light. The CO~$J= 8 \to 7$, $9 \to 8$ and $10 \to 9$ transitions have rest frequencies of 921.80, 1036.91 and 1151.99 GHz and half power beam widths (HPBWs) of 23, 20 and 19 arcseconds respectively. We use 
\begin{align*}
\Omega = \frac{\pi}{4\ln 2}\left(\mathrm{HPBW}\right)^2
\end{align*}
to compute the beam area. Finally, we use the average of the detected line intensities for comparison (Column 8 of Table 3 in Pon15). For this set of values the line fluxes are shown in Table~\ref{tab:IRDClines}. The second column shows the CO~$J= 8 \to 7$ line flux, while the third and fourth columns show the CO~$J= 9 \to 8 / 8 \to 7$ and CO~$J= 10 \to 9 / 8 \to 7$ line ratios respectively. The CO~$J= 10 \to 9$ line was not detected in any of the clumps so we use the upper limits adopted in Pon15. This means that the CO~$J= 10 \to 9 / 8 \to 7$ line ratio is an upper limit.

We also list in Table~\ref{tab:IRDClines} the predicted values from selected slow and fast shocks by multiplying the RADEX intensities (Figure~\ref{fig:intensities}) by $\Omega/4\pi$ for the appropriate beam areas. The predicted value of the CO~$J= 8 \to 7$ integrated intensity assumes the shock front faces the observer and fills the beam. We chose the models that give the closest CO $J= 8 \to 7$ line fluxes (Slow and Fast shock A in Table~\ref{tab:IRDClines}) as well as the shocks that give the closest CO~$J= 9 \to 8 / 8 \to 7$ line ratios (Slow and Fast shock B in Table~\ref{tab:IRDClines}).

Slow shocks A and B have shock velocities $v_s = 2 \kps$ and $3.5 \kps$ respectively. Fast shocks A and B have shock velocities $v_s = 3.5 \kps$ and $4.0 \kps$ respectively. All four shock models have preshock density $n_0 = 10^4 \, \rm{cm}^{-3}$ which agrees well with typical densities of IRDCs. Fast shock A is the only shock that doesn't overpredict the observed integrated intensities, but the predicted integrated intensities of the other three models could be reduced to match the observations if the shock doesn't fill the beam. Hence the observed line flux of CO~$J= 8 \to 7$ from these IRDCs can be explained by either fast or slow shocks.

The slow shocks generally fit the CO~$J= 9 \to 8 / 8 \to 7$ line ratios better than the fast shocks, with the line ratio from slow shock B (0.80) very close to the average of the ratio for all three clumps ($\sim 0.78$). The predicted line ratio from fast shock B (0.25) is the largest predicted from all the fast shocks modeled, which still underpredicts the observed values by $\sim 3-4$ times. Hence the high temperatures produced in slow shocks are necessary to explain both the CO~$J= 8 \to 7$ and $9 \to 8$ emission from these clumps. Combined with the inability of PDR models to explain these observations, we therefore suggest that a slow shock interpretation is favoured by the models in this paper.

\section{Conclusions}\label{ch:Conclusion}

The one dimensional time independent two-fluid MHD equations were numerically integrated to compare the structure of low-velocity fast and slow shocks in molecular clouds. \highlight{Our simplified model includes the effects of the major coolants found in molecular clouds and follows the abundances of chemicals affecting the production of H$_2$O in a simple chemical network. The solutions highlight important differences between fast and slow MHD shocks in molecular clouds.} These shocks show strong differences in their velocity and density structure because of the different driving pressures behind the shock fronts. Fast shocks are driven by magnetic pressure while slow shocks are driven by gas pressure. This means that the thickness of fast shocks is set by the long ion-neutral collision timescale, wheareas the thickness of slow shocks is set by the short neutral-ion collision timescale. The cooling timescale of the gas lies between these two and so peak temperatures in slow shocks are far higher than fast shocks of the same shock speed.

We showed that fast and slow shocks are observationally distinct and provided some example diagnostics. \highlight{For instance, low lying pure rotational lines of H$_2$ contribute negligibly to the cooling in fast shocks, whereas they produce significant radiation from the warm gas in slow shocks.} The non-LTE radiative transfer code RADEX was used to estimate line strengths of rotational transitions of $^{12}$CO, $^{13}$CO, ortho-H$_2$O and para-H$_2$O. The higher temperatures of slow shocks excite the high-$J$ transitions of $^{12}$CO more than in fast shocks. Line ratios near these transitions show strong differences between fast and slow shocks across the velocity range and therefore may be strong indicators of shock type. Anomalously strong high-$J$ CO lines have been observed in nearby infrared dark clouds \citep{pon_mid-j_2015}. The line ratios from these observations closely match slow shock predictions and are poorly fit by any of the fast shocks modeled here.

This suggests that simulations of MHD turbulence could gain observational predictions if the statistics of shock types were recorded. \highlight{If the mixture of shock families is found to be sensitive to turbulence parameters---such as the driving mode, the kind of feedback included, Mach number variations and self-gravity---then shock signatures become observational probes of the turbulence. Combined with tracers of star formation like young stellar objects, shock signatures could then shed light on the influence of supersonic MHD turbulence on the character of star formation.}

\section*{Acknowledgments}
The authors gratefully acknowledge discussions with Christoph Federrath and James Tocknell, and thank the referee for a prompt and stimulating report. This research was supported under Australian Research Council's Discovery Projects funding scheme (project number DP120101792). AL was supported by an Australian Postgraduate Award.

\bibliographystyle{mn2e}
\bibliography{bibliography}

\end{document}